\documentclass{cupconf}

\ifx\pdftexversion\undefined
\usepackage[dvips]{graphicx}
\usepackage{epsf}
\else
\usepackage[pdftex]{graphicx}
\usepackage{epstopdf}
\fi

\title[Dust Disk Lifetimes]{Observational Constraints on Dust Disk Lifetimes: Implications for Planet Formation}

\author[L. A. Hillenbrand]{Lynne A. Hillenbrand}
\affiliation{$^1$California Institute of Technology,
 MS 105-24, Pasadena, CA 91105, USA}

\pubyear{2006}
\volume{XXX}
\pagerange{XXXX}
\date{?? and in revised form ??}
\begin{document}

\maketitle

\begin{abstract}
Thus far our impressions regarding the evolutionary time scales for young circumstellar disks have been
based on small number statistics.  Over the past decade, however, in addition to precision
study of individual star/disk systems, substantial observational effort has been invested in obtaining less 
detailed data on large numbers of objects in young star clusters.  This has
resulted in a plethora of information now enabling statistical studies of disk evolutionary diagnostics.
Along an ordinate one can measure disk presence or strength through indicators such as ultraviolet/blue
excess or spectroscopic emission lines tracing accretion, infrared excess tracing dust, 
or millimeter flux measuring mass.  Along an abscissa one can track stellar age.   While
bulk trends in disk indicators versus age are evident, observational errors affecting both axes,
combined with systematic 
errors in our understanding of stellar ages, both cloud and bias any such trends.  
Thus detailed understanding of the physical processes involved in disk dissipation 
and of the relevant time scales remains elusive.
Nevertheless,  a clear effect in current data that is unlikely to be altered by data analysis 
improvements is the dispersion in disk lifetimes.  Inner accretion disks are traced by near-infrared
emission.  Moderating a generally declining trend in near-infared continuum excess and excess
frequency with age over $<$1 to 8$\pm4$ Myr, is the fact that 
a substantial fraction of rather young ($<$1 Myr old) stars apparently
have already lost their inner accretion disks while a significant number of rather old 
(8-16 Myr) stars apparently still retain inner accretion
disks.  The age at which evidence for inner accretion disks ceases  to be apparent 
for the vast majority ($\sim$90\%) of stars is in the range 3-8 Myr. 
More distant, terrestrial zone dust is traced by mid-infrared emission where
sufficient sensitivity and uniform data collection are only now being realized
with data return from the Spitzer Space Telescope.  Constraints on mid-disk dissipation
and disk clearing trends with radius are forthcoming.

\end{abstract}

\firstsection 
\section{Introduction}

A long standing paradigm for the formation of stars, and subsequently planets,
involves the rotating collapse of a molecular cloud core to form on a time scale of
$\sim$10$^5$ yr a central proto-star surrounded by an infalling envelope and 
accreting disk.  Typical ages of revealed young T Tauri and Herbig Ae/Be stars
are $\sim$10$^6$ yr. Gradual dispersal of 
the initially optically thick circumstellar material occurs in the early pre-main sequence phase
as the system evolves through the final stages of disk accretion, which can last 
 $\sim$10$^7$ yr or more in at least some well known cases (TW Hya, Hen 3-600, TWA 14 -- 
Muzerolle et al. 2000, 2001 and Alencar \& Batalha 2002; PDS 66 -- Mamajek et al. 2002;
ECha J0843.3-7905 -- Lawson et al. 2002; St 34 -- White \& Hillenbrand 2005).

Physical processes occurring in younger disks include viscous accretion onto the
central star, mass loss due to outflow, irradiation by the central star, ablation due to
the stellar wind, turbulent mixing of material, stratification,
and gradual settling of the dust towards the disk mid-plane --
this last process a critical and limiting step in the path towards planet formation 
in the standard core accretion model (e.g. Weidenschilling et al. 1997, 2000; Pollack et al. 1996).
The total disk mass decreases and the dust:gas mass ratio,
assumed at least initially to be in the interstellar ratio, changes  
with time due to a combination of the above effects.  Similarly, the
dust particules are assumed interstellar-like in their composition and 
structure.  Of particular interest here is the expected loss of dust opacity 
due to assembly of small particles into larger bodies that might later 
be known as planetesimals.  For solar-type stars, the ultimate result in at least 
10\% and perhaps as many as 50\% of cases is a mature solar system (see Marcy, this volume).

In parallel with the discovery and study of exo-solar planets and planetary systems
over the last decade (the topic of this conference),  
we have had dramatic observational confirmation in this same time period of the 
basic paradigm for star formation as briefly outlined above.  Direct images and 
interferometric observations which spatially resolve 
young circumstellar disks at optical, near-infrared, and millimeter wavelengths  have 
become common, though are far from ubiquitous.  When combined with measured spectral energy
distributions, such spatially resolved data are valuable for breaking model
degeneracies and thus improving our understanding of source geometry and dust characteristics.

Rough correlation of the spatially resolved and SED appearances of a source, which indicate
{\it circumstellar} status, with 
{\it stellar}  evolutionary state, or age, has long been advocated (e.g.  Lada 1987).
However, it remains unclear whether the established sequence of circumstellar
evolutionary states corresponds directly with source age. White \& Hillenbrand (2004)
argue for the Class I/II stages that this is not necessarily the case given the similarities
in the stellar photospheric  and accretion properties of Class I and II stars
as inferred from high dispersion spectroscopy of a large sample in Taurus-Auriga.   
Likewise, Kenyon \& Hartmann 1995 discuss the Class II/III distribution in the HR diagram,
which is indistinguishably  intermingled and therefore suggestive of similar ages.  
Because of uncertainties in age assignments,  particularly for the most enshrouded
sources which typically do not have ages estimated independent of their circumstellar 
characteristics, the time scales
associated with the dispersal of circumstellar material and the formation of planets
are only vaguely constrained at best.

\begin{figure}
\begin{center}
\vskip -1.5truein
\hskip -0.2truein
\includegraphics[width=0.95\textwidth]{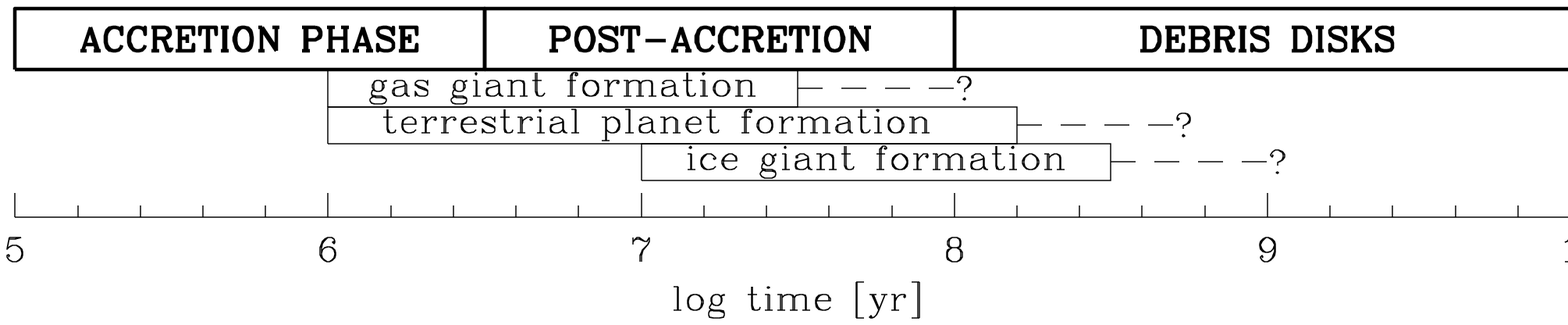}
\vskip -3.5truein
\includegraphics[height=113pt]{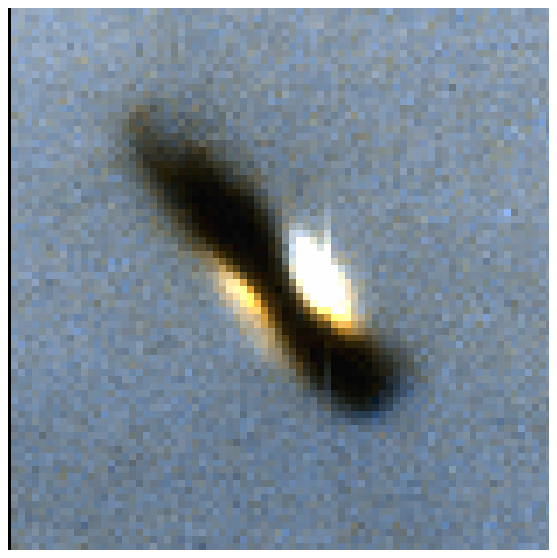}
\includegraphics[height=113pt]{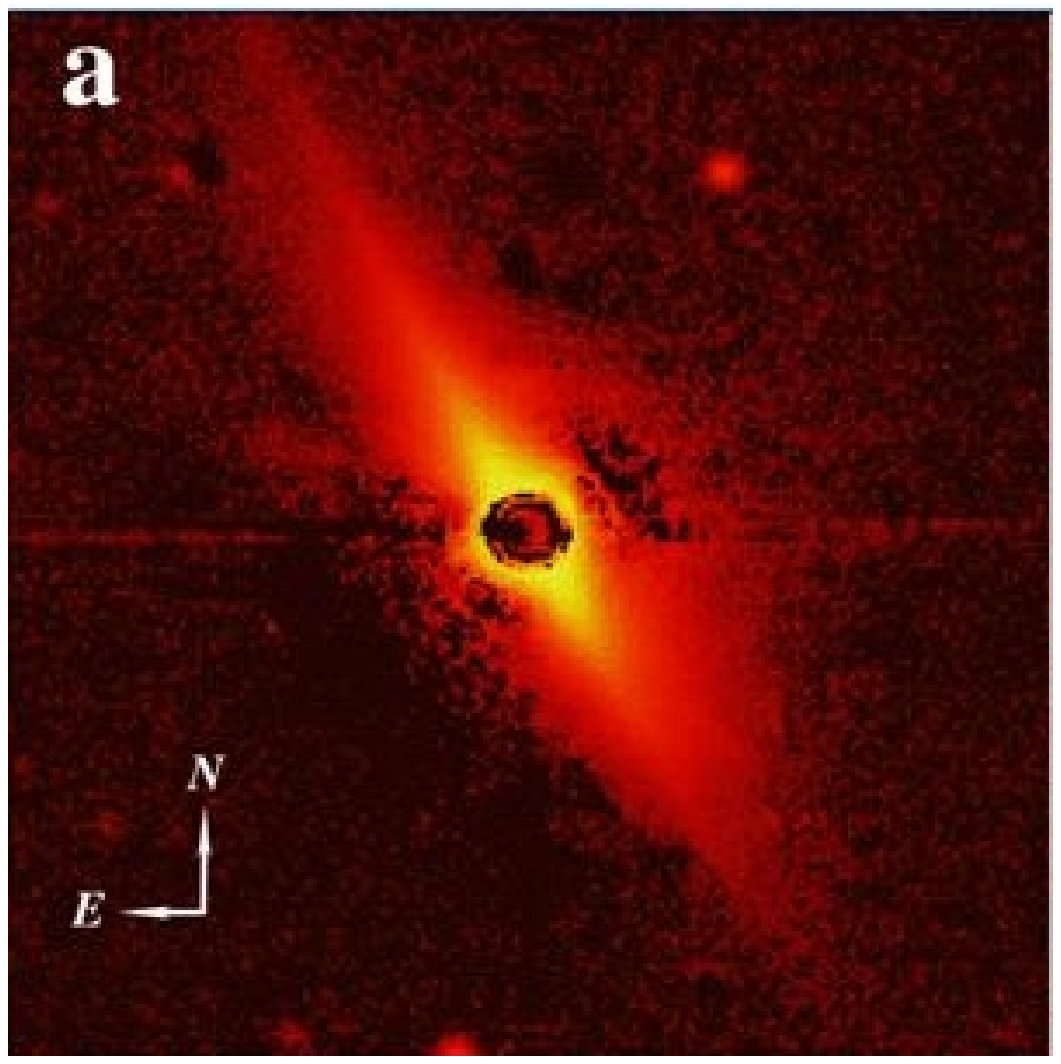}
\includegraphics[height=113pt]{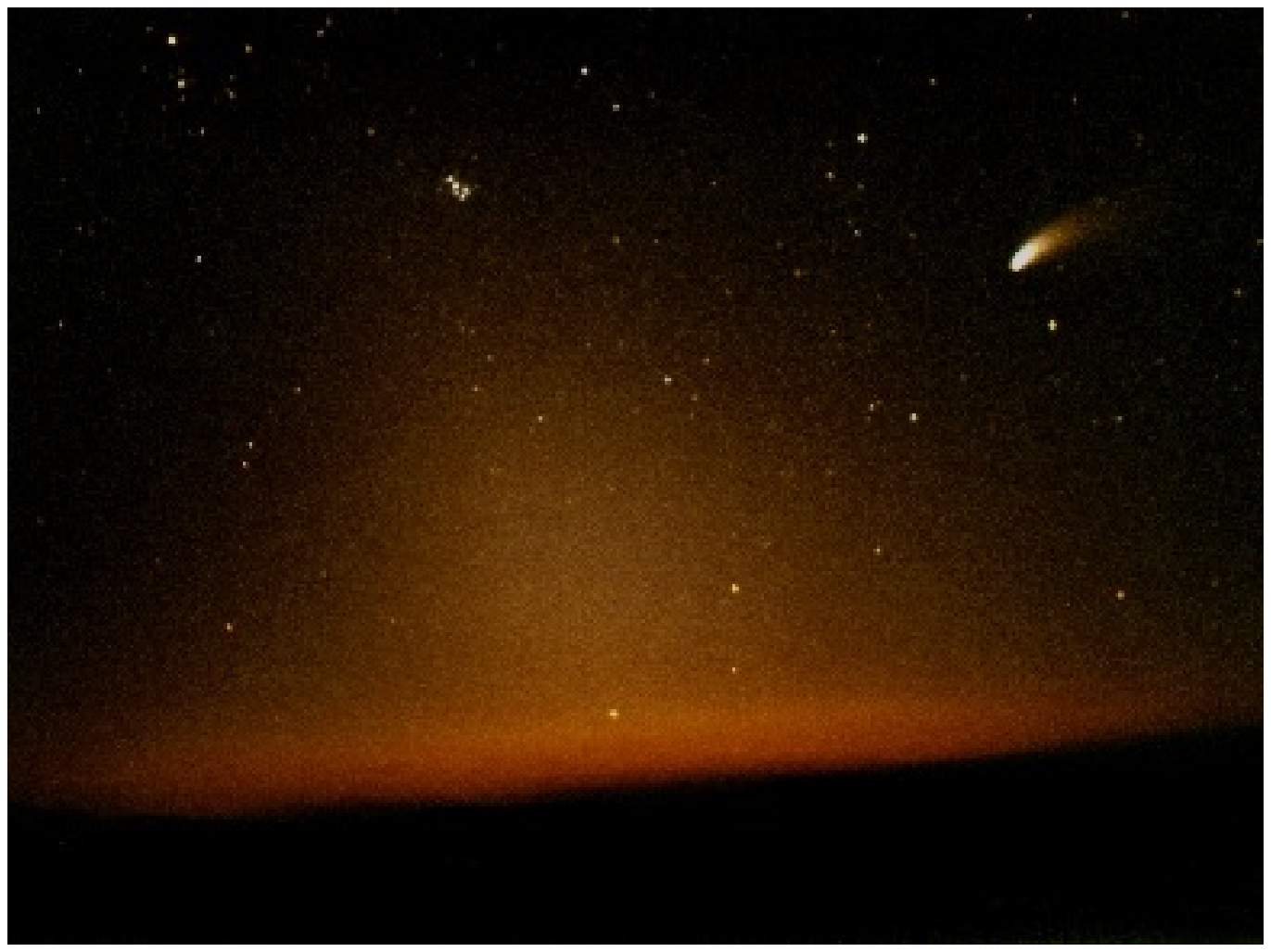}
\vskip 1.5truein
 \vspace{0cm}
\caption{
{\it Images of disks at various evolutionary stages
scaled to a time line showing our general understanding 
of the basic phenomena.}
Data are courtesy of
J. Stauffer and B. Patten (left panel, Ori 114-426 optically thick
``silhouette disk'' as imaged with HST/WFPC), 
Kalas \& Jewitt 1995 (middle panel, $\beta$ Pic as imaged by a ground-based
coronagraph), and
P. Kalas (right panel, our own zodiacal dust disk along with a comet, as
photographed from Calar Alto).}
\label{hillenbrand_fig1}
\end{center}
\end{figure}

How, then, do we catalog young circumstellar disks and characterize their evolution?
Disk diagnostics  come in two forms:  those that trace
the dust and those that trace the gas.  Dust implies small particles 
with typical tracers sensitive to sizes less than 1 mm.  
These include continuum spectral energy distributions over
several decades in wavelength, solid state spectroscopic features
in the mid-infrared, and direct images measuring either thermal emission
at long wavelengths (mid-infrared through millimeter) or scattered light at
shorter wavelengths (optical and near-infrared).  Gas tracers should reveal the
bulk of the mass, at early stages more than 99\% of the total mass if 
interstellar abundances can be assumed.  Sensitive gas observations 
of young circumstellar disks are, however,
limited thus far,  restricted to trace species, and dominated by upper limits.
Yet recent observations of CO, H$_2$, and H$_2$O seem promising for
characterization of terrestrial zone gas.
Najita  (this volume) presents our knowledge of gas disk evolution in detail.

 In addition to
academic interest in disk dissipation mechanisms, the main motivation for
understanding disk evolution time scales is the relation to planet formation.
It seems prudent then to begin with a summary of the {\it capacity} of young disks
to form planets.   We will then continue with methods for assessing
the {\it probability} that young disks do indeed form planets.

\section{The Potential for Planet Formation in Young Circumstellar Disks}

What are the initial conditions in young disks and what is the likelihood 
that they are in fact proto-planetary? The raw material of planetary embryos, 
Earth-like rocks, and Jupiter-like gas giants is indeed abundant, if not ubiquitous,
 in young disks.  But whether any individual disk {\it will} form planets 
is of course unknowable.  What we can say is that many of the disks we observe are at least
{\it capable} of forming planetary systems similar to our own, as evidenced from measured
disk sizes, masses, and composition/chemistry.  However, as detailed below, the {\it mean}
disk properties are not yet known due to sensitivity limitations
and therefore comparisons to our own proto-solar system based on
existing data may be biased.

Disks around young stars were spatially resolved for the first time at millimeter
wavelengths (e.g. Sargent \& Beckwith 1987) which measure cold dust and gas 
in the outer disk regions.  Unequal axial ratios, combined with implied dust
masses large enough that the central stars should not be optically visible 
if the dust geometry is spherically symmetric, stood as the strongest evidence for close 
to a decade of disks surrounding young  stars.  Further,
kinematic models of spatially resolved CO emission demonstrated consistency with
Keplerian rotation  (e.g.  Koerner et al. 1993; Mannings et al. 1997; 
Simon et al. 2000; Qi et al. 2003).

Continued interferometric work (e.g. Lay et al. 1994,
Dutrey et al. 1996, Duvert et al. 2000; Kitamura et al. 2002;
Qi et al. 2003, Semenov et al. 2005).
suggested that disk diameters -- in instances where spatially resolved, 
as opposed to point-like, images are obtained -- range from 
$\sim$70-700 AU and are even as large as $\sim$2000 AU in some cases.  
These disk size estimates are consistent with those inferred
from optical/near-infrared scattered light or silhouette images 
(e.g. McCaughrean \& O'Dell 1996; Padgett et al 1999; Bally et al. 2000), 
and in the typical case are comparable to or larger than the orbit of the outermost
gas giant in our solar system, Neptune.  Surface density
profiles, e.g. simple power-laws with $\Sigma(r) \propto r^{-p}$ or viscous disk
``similarity solutions'' with $\Sigma(r) \propto r^{-p} e^{-r^{(2-p)}}$, 
have suggested a wide range in the value of p (0-1.5 for the power-law case).  

Disk masses are derived from optically thin millimeter flux and an
adopted opacity-wavelength relationship which leads to uncertainties of 
factors of 5-10 in disk masses.  Under common assumptions the calculated dust masses 
range from 10$^{-4.5}$ to 10$^{-3}$ M$_\odot$ (e.g. Beckwith et al. 1990).
Making the further assumption that the dust:gas ratio by mass is 
unaltered from the canonical interstellar value of 1:100, 
total disk masses average around 0.02 M$_\odot$, or about the
Minimum Mass Solar Nebula (Kusaka et al. 1970;
Weidenschilling 1977), the reconstitution
of present-day solar system mass and composition to solar consistency.
It should be stressed that {\it detection at all of millimeter flux} is made amidst an increasing
number of upper limits measured for stars with other indicators of disks at shorter wavelengths, 
and so the true ``mean mass" is even lower than that quoted above.

The composition of both young primordial and older debris disks has been
shown to resemble that of solar system comets. Ground-based 
10 and 20 $\mu$m work on samples of brighter sources 
(e.g. Hanner et al. 1995, 1998; Sitko et al. 1999; 
van Boekel et al 2003; Kessler-Silacci et al 2005) and
especially 
ISO 2-30 $\mu$m spectroscopy (e.g. Meeus et al. 2001; Bouwman et al. 2001)
have revealed an impressive suite of solid state (and PAH) dust features.
Mineralogical details of the dust are modeled on a case-by-case 
basis due to cosmic variance, but the mean composition 
appears to be $\sim$70-80\% amorphous magnesium-rich olivines,
$\sim$1-10\% crystalline forsterite, $\sim$10-15\% carbons, 
$\sim$3-5\% irons, and other trace components such as silicas.
In particular, crystallinity is advocated in $\sim$10\% of sources.

In summary, the {\it observed} sizes, masses, and chemical composition
of young disks are all consistent with solar nebula estimates.  This is a weak
statement, however, since the {\it mean disk properties} are biased at present
by detection limits and selection effects.

\section{Questions Concerning ``Primordial" Dust Disk Evolution}

The term ``primordial" is used in reference to disks that are remnants of
the star formation process.  As outlined above, such disks are composed of
dust and gas which participated in the gravitational collapse that formed the star
and now  comprise  the raw materials for
the formation of planets.   The size, mass, and composition parameters of known
young primordial disks are consistent with those estimated for the proto-solar system
disk. Terrestrial planets and the rocky cores of giant planets
originate from disk dust while the gaseous envelopes of giant planets
originate with the disk gas.  Primordial disks are in the process of dissipating
through either planet formation or one of the other disk dispersal mechanisms
mentioned earlier.  

It is instructive to point out that primordial disks are physically
distinguished from the so-called ``debris" disks, which are secondary
rather than primordial.  These are gas-poor disks, comprised of dust
which is regenerated during and subsequent to the growth of planets as
the large/massive bodies incite collisions amongst smaller bodies to re-form dust. 
Debris disks, like primordial disks, are in the process of dissipating, though via a
different mechanism.   Rather than sticking collisions which result in smaller particles
growing to become larger particles (and eventually becoming undetectable via
thermal infrared radiation), debris disk particles experience shattering collisions
and gradually grind themselves down to the point at which grains are efficiently removed
from the system via effects such as Poynting-Robertson drag and stellar winds.  However,
new dust is continually being generated in the cascade generated by collisions 
between the larger bodies and the dissipating evolution is punctuated by the infusion of new
material in the debris cascade.  

The collisional history  in the inner solar system, 
due to the influence of the outer giant planets on such debris, is well-documented
in the cratering records on the Moon and Mars.  These records indicate to some degree the evolution
of the cratering rate and the large body size distribution with time.  We have no firm record of the
dust evolution in solar system, but even today there is
``debris dust" found/assumed in the Asteroid and Kuiper belt
regions.   Because of the strong theoretical connection between debris dust and planetary
perturbers, there is much interest in the debris belts seen around stars other than the Sun
(see Meyer, this volume) whose evolution we can study by investigating samples of
different age.

 Here I focus on the properties and evolution of dust in {\it  primordial disks}.  For any
 given disk, the dust mass is expected to decrease with time throughout the duration of
 the planet-building process, perhaps over tens of Myr.  Then, if planets have
 successfully formed, the dust mass increases at the on-set 
 of the debris disk phase before slowly declining again with time over many Gyr.
 
To understand the process of planet formation we must understand how quantities such as
initial disk size and radial/vertical structure,  initial disk mass and mass surface density, and the 
initial disk composition and chemistry all evolve with time and, further,
 the relative importance of various disk dispersal mechanisms (e.g. accretion, 
ablation, grain growth as mentioned above).  Over what time scales are
dust (and gas)  detectable and how does the mass ratio of dust:gas evolve?  
What physical parameters determine disk longevity?
What is the frequency of different end states, in particular of planetary configurations?   
Most important for understanding the rarity or commonality of the formation 
of our own solar system,
what is the mean and the dispersion in all of the above distributions?

As we continue to develop the tools for answering these questions we can also consider
several pertinent ``second parameter" issues. 
One category of these relate to properties of the central star. Are there correlations 
in initial disk properties or disk evolution diagnostics with stellar properties such as
the radiation field (particularly x-ray and ultraviolet output), stellar mass, 
or system metallicity, all of which may have important effects  on disk structure and chemistry?  
A second category of second parameters are those related to disk physics effects.
How does disk accretion history, in particular poorly understood outburst phenomena
such as FU Ori or EXOr type events affect disk evolution? Thirdly, what is the role of
environment?  Multiplicity in the form of binary, triple, or quadruple systems can
influence disk evolution when the  companions are within or just exterior to the disk.
Clustered versus isolated star forming environments in which effects such as increased
ionization or photo-evaporation of disk material by massive stars, dynamical effects 
due to high stellar density, or the mechanical effects of multiple jets/outflows, could be important
for disk evolution.   Consequently, understanding of multiplicity statistics in the form of
frequency and orbital parameters and clustering statistics in the form of spatial density
and luminosity function
is important for our appreciate of the range of plausible disk evolutionary paths.

In summary, there are many parameters considered potentially influential in the disk
evolution process.  The only way to probe effectively disk evolution and its many
dependencies is through the assembly of sufficient statistics over the appropriate range
of ages and ``second parameter" conditions.  This is a tall order indeed, but a road down
which we have at least started.

\section{Enough Questions - What do we Know and How do we Know it?}

The disk dispersal time or disk lifetime is often asserted in the literature 
as ``about 10 Myr".  This estimate is
certainly good to an order of magnitude, but the justification for this number, or any other
specific number, is weak at best given the data in hand. 
 {\it Some} inner dust/gas disks have disappeared
within 1 Myr, by the time the star becomes optically visible.  {\it Some} inner dust/gas disks last
at least 10 Myr.   In {\it at least one case}, that of our own solar system, the need has
been expressed by some theorists for the gas disk to survive 100 Myr  or longer,
in order to form the outermost gas giants.    As astronomers, we want to understand
the mean and the dispersion in the lifetime time of young primordial disks,
both dust and gas.   In this section
I will review disk diagnostics, appropriate subject samples, and the difficulties involved
in assessing stellar ages.   In the next section I  proceed to summarize what is known about disk evolutionary trends.

\subsection{Disk Diagnostics}

To look for evidence of disk evolution in action we need to consider carefully the 
diagnostic potential of any particular observable.   Many are available.  However, the
information obtained varies widely between different tracers of disk evolution.  This is due in part
to variation in observational sensitivity, for example as a function of wavelength,
 and in part to the varying efficacy of different disk tracers.   In addition,
the precision and accuracy of stellar ages -- that other, often under-scrutinized  or even
ignored axis in any disk evolution diagram -- needs to be critically assessed. 

Deferring sensitivity considerations for the time being,
what can we hope to measure as a function of stellar age? 
Resolved disk images as discussed in my introductory comments certainly 
have led to a wider appreciation of the convincing case for primordial
``proto-planetary" disks.  In fact, it is not an over-statement to say that
the stunning images from ground-based interferometers (millimeter) and from 
the Hubble Space Telescope (optical and near-infrared) were responsible 
for transforming the field of star formation from a following of dedicated and knowledgeable 
disciples to high profile science.    However, the reality is that few such 
spatially resolved images exist at present.   Study of most young disk systems
relies, for the most part,
on so-called indirect measurements such as broadband photometry and
high resolution optical or near-infrared spectroscopy.

Table 1 details several properties of quantitative interest for young circumstellar disks
and the observational diagnostics used to measure them.  These are generalized properties and
each can be broken down into a more detailed set of specific physical characteristics.  There is
an increasingly large literature on these topics and I list only a few example studies.  In most
categories there is some, but limited, evidence for at least modest evolution from primordial disk 
conditions.  Conclusions in the area of evolution are typically based on samples 
of young disks ranging from small (a few) to moderate (tens to a few hundred) in size.

\begin{table}
\scriptsize
  \centering
 \caption{Dust disk properties measurable as a function of stellar age }
  \begin{tabular}{@{} lll @{}}
   \hline
  Property  & Observational Diagnostic & Example Study \\ 
    \hline
    Disk geometry & Interferometry; SED modelling & Eisner et al 2004, 2005\\ 
    Mean excess in SED; disk fraction & Broad-band photometry & Hillenbrand et al. 2006; Mamajek et al 2004\\ 
    Accretion rate on to star& Ultraviolet/optical spectrophotometry & Muzerolle et al 2000; White \& Hillenbrand 2004\\ 
    Dust mass & Millimeter/sub-millimeter photometry  & Carpenter et al. 2005; Wyatt et al. 2003\\ 
    Dust mineralogy; size distribution & Mid-infrared spectroscopy & Kessler-Silacci et al 2005; van Boekel et al 2005\\
      
  \end{tabular}

  \label{tab:tab1}
\end{table}

As mentioned above,
the focus of my discussion will be on dust disk diagnostics.
In particular I will focus on disk detection as revealed through infrared excesses,
observed emission in excess of that expected from a stellar photosphere.   Various levels 
of sophistication may be employed in the application of this technique,
ranging from fully assembled spectral energy distributions covering several
decades in wavelength to two-color diagrams which cover only a limited portion of the
excess spectrum to statistical study of the disk fraction (frequency of objects in a given age bin
with convincing evidence of a disk) or mean excess (magnitude or strength
of the excess).  Full spectral energy distributions covering ultraviolet to millimeter wavelengths
 have been available for only small samples of well-studied
young stellar objects, making statistics difficult to assemble.  Two-color diagrams are
widely available, enabling statistical studies, but more difficult
to interpret without detailed knowledge of  1) more of the spectral energy distribution 
2) the intrinsic spectral energy distribution in the absence of reddening which can be
prevalent towards young star forming regions
 and 3) the properties of the underlying star.   Disk fraction and mean excess techniques
 account for both reddening and intrinsic stellar colors, but are based on partial
 spectral energy distributions.   The discussion below will focus on these last techniques.

Data for infrared excess investigations are available most
abundantly at near-infrared wavelengths, due to several decades of ground-based
work combined with  the large and uniform 2MASS photometric database at
1.2, 1.6, 2.2 $\mu$m now available.   Ground-based work at mid-infrared wavelengths has been more
limited in both scope and sensitivity;  previous space-based platforms were revolutionary
at the time, but somewhat similarly limited in sensitivity (IRAS) and scope (ISO).  Thus our
understanding of disk statistics in the 3-100 $\mu$m wavelength regime is not as well 
developed as in the 1-2$\mu$m regime.  The Spitzer Telescope is currently accumulating 
sensitive data between 3 and 70 $\mu$m enabling the construction of mid-infrared spectral energy
 distributions.  These observations focus on 
many of the historically favored objects, though blind imaging surveys
of star forming regions and young open clusters are also being conducted 
and will provide needed statistics over the next few years.

\subsection{Stellar Samples }

Once a technique is adopted, a sample must be chosen.
In order to establish trends, robust, complete, and unbiased samples must be established
over an appropriate range of ages.  For the problem at hand this includes
the youngest revealed protostars through
the ages characteristic of star-forming regions still associated
with molecular gas ($<$1-2 Myr), and continuing to the entire period of  terrestrial and gas giant
planet formation (thought to be $\sim$100 Myr for our own solar system), as depicted
schematically in Figure 1.

Young star clusters would appear ideal for these sorts of studies because they provide the 
needed statistics.  Furthermore, clusters have attractive attributes such as the relatively uniform 
distance, age, and chemical composition of their members, all of which minimize analysis
complication.  Young star clusters can therefore, in principle,
provide the samples required to compare disk properties such as the mean and dispersion 
in disk lifetimes as a function of stellar mass (within a cluster) 
and as a function of stellar age or chemical composition (between clusters). 
However, careful investigation reveals that the young star samples
identified to date are lacking with respect to some important issues.

First, known targets for investigations of disk evolution can be segregated into the following
four coarse age groups:  $<$1 Myr (embedded or partially embedded star forming regions), 
1-3 Myr (optically revealed stellar populations still associated with
molecular gas), 5-15 Myr (association members in gas-poor 
``fossil" star-forming regions), and finally the punctuated ages
(55 Myr, 90 Myr, 120 Myr, and 600 Myr)  of the nearest populous open clusters.

Of concern is that the age distribution of known samples of young stars over the
1-100 Myr age range of interest for disk evolution is not uniform.  Ample numbers (many thousands)
of young stars associated with regions of recent star formation have been identified through
surveys  of molecular cloud complexes.  Because of the intense focus on stellar census data
for these young regions, such stars dominate the total numbers and 
thus bias the available statistics in the overall
young star age distribution towards  the 1-3 Myr youngest age category.   

The statistics decline dramatically at ages older than about 5 Myr and out to
about 50 Myr, due to a lack of large identified samples with known ages in this ``young intermediate" 
age  regime.  There are no open clusters or large associations within 150 pc or so of the Sun
 in the 5-50 Myr age range save the Sco OB-2 association at the upper distance and lower age limit. 
 Field stars 5-50 Myr old  are extremely hard to identify since they
stand out from much older field star populations only with detailed 
observations (not, e.g., in wide-field photometric surveys).
They may be revealed through signatures of youth such as common proper motion
with kinematically young groups, enhanced
Li I absorption, Ca II H\&K core emission, and X-ray activity.
In fact, finding stars in this age range should be relatively easy due to our 
circumstance in the Galaxy near a ring of moderately recent star formation 
(``Gould's Belt").  Yet within 150 pc or so, current samples of 5-50 Myr old stars
number only in the few tens, consisting of members of the TW Hya, Beta Pic, Eta Cha,
and Tuc/Hor moving groups.  Continued
correlation of large-scale kinematic and activity databases with sufficient
spectroscopic following is beginning to address this 
deficiency.   However, the present lack of ample numbers of young stars 
in the 5-50 Myr age range serves to increase the error  bars in disk evolution diagnostic
statistics right where all the most interesting  ``action" of disk evolution may be taking place.

At ages older than 50 Myr there are again ample samples due to the proximity of several well-studied
near-by open clusters.  Specifically, the IC2602 / IC2391 pair,  Alpha Per, Pleiades, and Hyades
clusters, all within 200 pc and well-studied, are benchmark points in any evolutionary diagram
involving either stellar or circumstellar properties.

\subsection{Stellar Ages}

\begin{figure}
\begin{center}
\includegraphics[width=1.05\textwidth,clip]{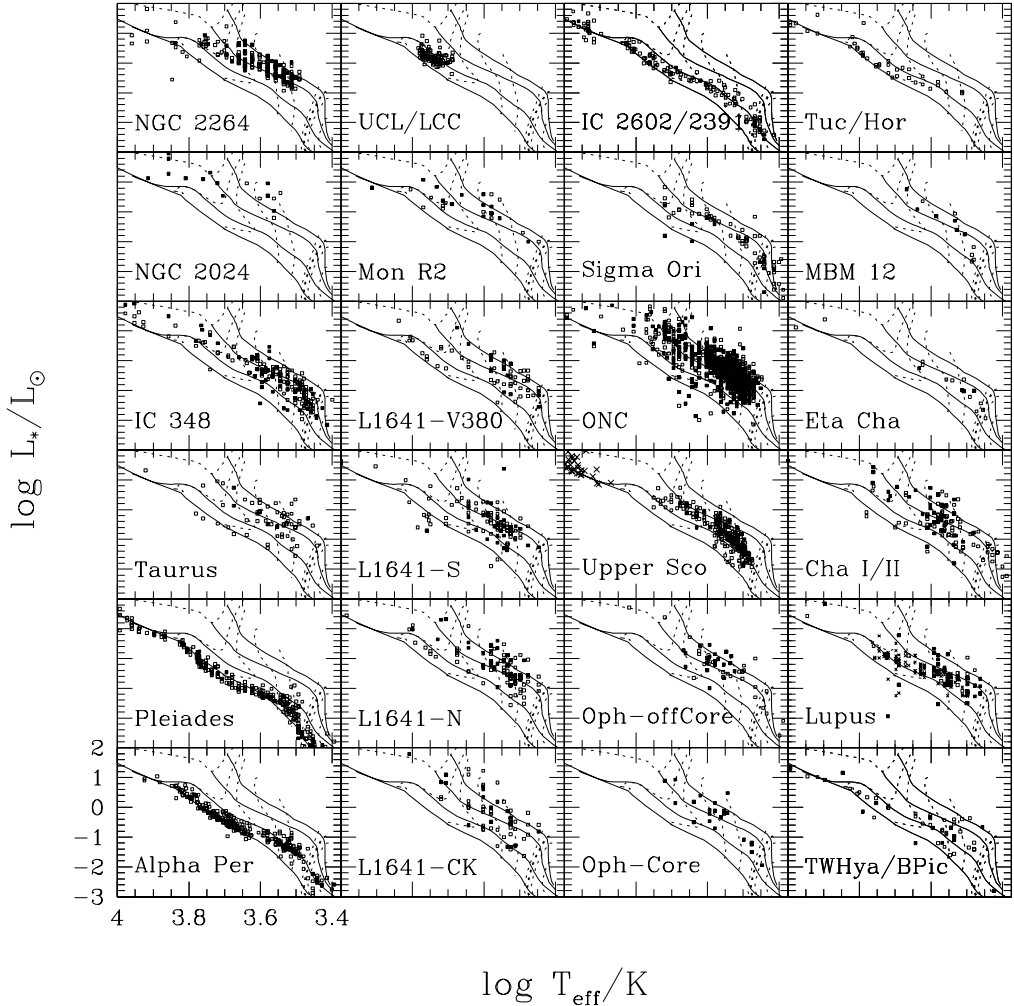}
\caption{ HRDs for well-studied star-forming regions and young clusters.
Data were placed using the temperature scale, color scale, and
bolometric corrections described in Hillenbrand \& White (2004) and a wide
variety of literature for the fundamental data.
Pre-main sequence evolutionary calculations are those of D'Antona \& Mazzitelli
(1997, 1998) for isochrones of 0.1, 1.0, 10, and 100 Myr (solid lines) and
masses 0.03, 0.06, 0.08, 0.1, 0.2, 0.4, 0.8, 1.5, and 3.0 M$_\odot$
(dashed lines).  
}
\label{default}
\end{center}
\end{figure}

A comprehensive discussion of stellar ages is beyond the scope of this review.  Suffice it
to say that there are large number of age diagnostics, most of which are poorly calibrated in the
young pre-main sequence age range of interest here.  The most commonly used measure of
stellar age in the $<$1-30 Myr age range is location in the Hertzsprung-Russell (HR) diagram compared
to theoretical predictions of luminosity and temperature evolution as a function of time.
HR diagrams are shown  in Figure 2 for a number of current and recently star forming regions as 
well  as young open clusters in the solar neighborhood.  

HR diagrams can be used to infer a mean age and an apparent age dispersion as a function
of effective temperature, for each cluster.
One issue to consider is whether the age spreads one would infer for
stars in young star clusters from their observed luminosity spreads indeed correspond to age ranges
rather than observational errors, the default assumption.  Luminosity spreads do decrease
with time (e.g. consider the Orion Nebula Cluster vs the Alpha Per Cluster in Figure 2). 
However, the errors in converting from observables to  luminosity are  the largest in 
the young pre-main sequence phase, just where the apparent luminosity spreads 
are the largest.  Thus the conversion from luminosity to age, and the implied age spreads,
are confusing.   Age spreads, or lack thereof, are important to understand 
for the purposes of studying evolutionary diagrams since one needs either to
consider all stars in a single cluster to have the mean age of
the apparent distribution, or to assess ages individually and adopt an age for 
each star.  Because this issue has
not been satisfactorily addressed at the young ages of interest here, the evolutionary time 
scales for young circumstellar disks thus have large \emph{random} uncertainties depending on whether
potentially real age spreads are accounted for in the analysis or not.

Another caution  is that theoretical pre-main sequence evolutionary calculations, on which age estimates
from the HR diagram rely, have significant uncertainties in their predictions. First,
there is variation between various theory groups of 20-100\%  over certain mass and age ranges
(see comprehensive discussion in Baraffe et al. 2002).
Second, pre-main sequence calculations 
thus far do not favor well in comparison to observational constraints.  Specifically,
they collectively {\it under-predict stellar masses by 30-50\%} (Hillenbrand \&  White 2004).  
Further, they {\it under-predict low-mass stellar ages by 30-100\%} compared to lithium-depletion boundary estimates and {\it over-predict high-mass stellar ages by 20-100\%} compared to 
post-main sequence evolutionary calculations.  
 Because of this lack of theoretical validation
of the age calibration of pre-main sequence isochrones, the evolutionary time scales for young
circumstellar disks thus have large \emph{systematic} uncertainties.

\section{Disk Evolution}

I describe now the observational constraints on the evolution of 
potentially proto-planetary disks through the disk clearing phase.
As already emphasized I will focus on  dust disk evolution, mentioning gas where
it should not be forgotten, but not discussing gas in any detail.
Three spatial regimes in the disk are considered:  inner disk dissipation traced by
near-infrared continuum data, mid-disk dissipation traced by mid-infrared data,
and outer disk dissipation traced by millimeter wavelength data.

\subsection{Inner Disk Dissipation}

There is a well-demonstrated empirical connection between accretion and outflow 
diagnostics measured by high dispersion optical spectroscopy which probes the
kinematics of warm gas in the vicinity of young stars (e.g. Hartigan et al.  1995; 
White \& Hillenbrand 2004).   A similar empirical connection 
(e.g. Hartigan et al.  1990, Kenyon \& Hartmann 1995)
exists between the same spectroscopic emission lines 
and the blue continuum excess measured as spectroscopic veiling, both 
signatures of accretion directly onto the star, and photometric
near-infrared (1-3 $\mu$m) continuum flux excess arising in the innermost ($<$0.05-0.1 AU),
and thus hottest, disk regions.   These correlations affirm the basic connection 
between accretion from a disk and ejection in an outflow. 


Furthermore, both the spectroscopic signatures of accretion and the  near-infrared excess
are separately demonstrated to correlate inversely with stellar age, over small age ranges.
Detailed modeling of the accretion temperature, density,  velocity,
and geometric structure is required to convert emission line strengths and profiles
into mass accretion rates.  More common than emission line profile studies
is the measurement from high dispersion spectroscopy of continuum veiling
which can also be converted to a mass accretion rate after making assumptions about
the bolometric correction to derive a total accretion luminosity, and about the infall geometry.

Treatments of the trends in the accretion rate with age have been presented by 
Muzerolle et al. 2000 and Calvet et al. 2005a.  
At least several stars appear to show measurable accretion signatures beyond 10 Myr.
Existing trends have been inferred by considering the individually derived ages of stars based on 
the HR diagram.  They are thus subject to the criticism that age spreads in individual
clusters such as Taurus, Chamaeleon, or the TW Hydra association may be overestimated and
that comparisons between the mean accretion rates and mean stellar ages in each cluster
cluster may be more appropriate.   Similar criticisms are also levied below against treatments 
of near-infrared excess behavior with age.
  
For the near-infrared continuum analysis, we utilize measured flux
above expected photospheric values to infer disk presence.  
Increasingly complex inner disk geometries have been 
advanced over the last decade  (e.g. Mahdavi \& Kenyon 1998), which complicates the
expectations regarding the magnitude of a near-infrared excess given constant other 
parameters for the star and the disk.
In the analysis discussed here we do not consider such geometric complications
and assess simply whether there is, or is not, evidence for disk emission at
near-infrared wavelengths for our sample.   

In calculating the color excess due to the disk, one must make two corrections from 
observed colors.  First, it is necessary to derive and subtract  the 
contribution from foreground or large scale circumstellar extinction.   Second, from 
the remaining color, a correction for the underlying stellar photosphere 
is performed in order to arrive at the intrinsic color excess due to the disk.   
In formulaic terms, using H-K color as an example, the disk excess is quantified as
$\Delta(H-K) = (H-K)_{observed} - (H-K)_{reddening} - (H-K)_{photosphere}$.  Similar
indices can be derived for J-K or K-L colors which also probe inner disk regions though
sense dust at slightly different temperatures.
In order to effect the above extinction and photospheric corrections,
and hence assess intrinsic color excesses, several
different sets of information are required:
1) a spectral type, for intrinsic stellar color and bolometric correction
determination, 
2) optical photometry, for dereddening and locating stars
on the HR diagram, assuming known distance, 
and 3) infrared photometry, for measurement of disk ``strength".

It should be borne in mind that disk strength, quantified as above from measurement
of the absolute value of the infrared excess, is still a relative quantity. 
For any given star/disk system the infrared excess is affected by both 
stellar properties (mass, radius) and disk properties (accretion rate, 
inclination, geometry).  Meyer et al. 1997 and Hillenbrand et al. 1998 
(both in collaboration with Calvet) provide detailed discussions
of these dependencies apropos near-infrared excesses.  The effects of stellar
and disk parameters on overall spectral energy distributions are discussed 
more comprehensively by D'Alessio et al. (1999).

\begin{figure}
\includegraphics[width=0.45\textwidth,clip]{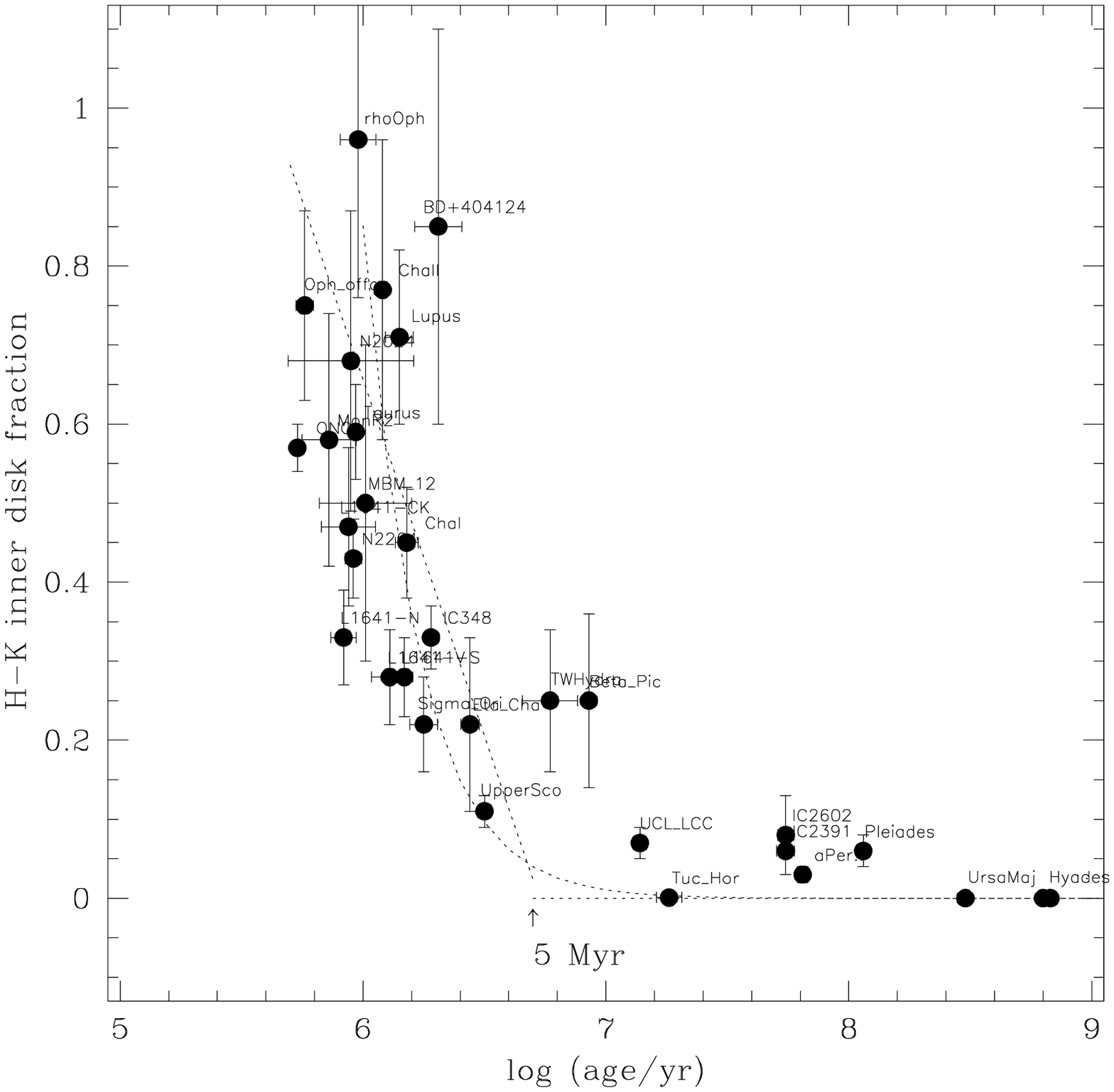}
\includegraphics[width=0.45\textwidth,clip]{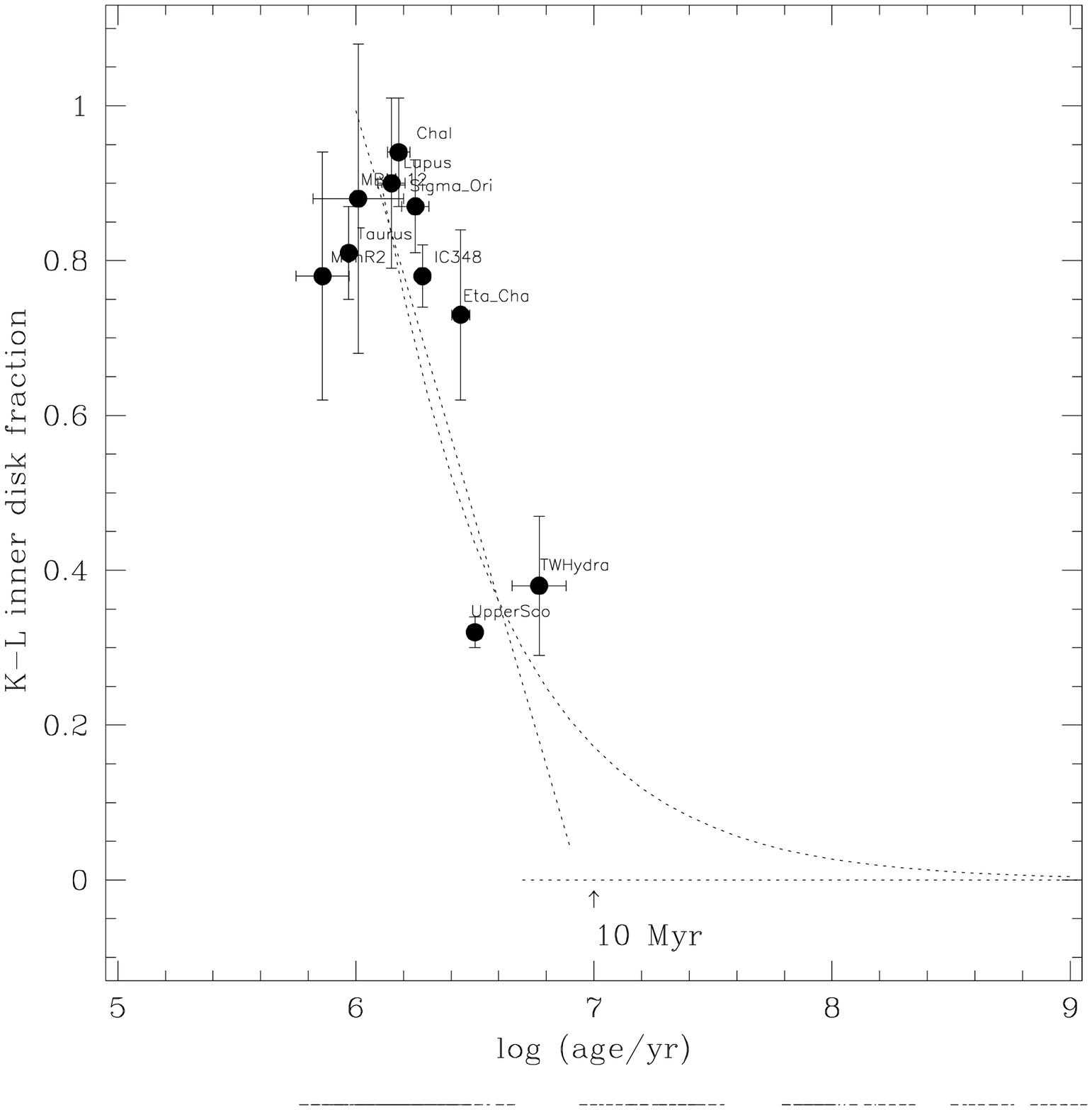}
\caption{
  {\it Inner accretion disk fraction vs. stellar age} inferred from
  H-K excess (left panel) and K-L excess (right panel) measurements,
  binned by cluster or association.  
  All young stars which we are able to locate in the HR diagram based
  on information in the literature (about 3500) and having inferred masses 
  0.3-1.0 M$_\odot$ are included in this figure. Individual
  clusters are treated as units of single age corresponding
  to the median age inferred from the HR diagram.  
  A cut of $\Delta$(H-K) $>$ 0.05 mag is used to define a disk.
  Standard deviation of the mean (abscissa) and Poisson (ordinate)
  error bars are shown.  The linear and exponential fits were derived 
  for ages $<$30 Myr; the linear fit has negative slope close to unity
  with rms 0.3. }
\end{figure}

Now what about that pesky other axis of stellar age?  Instead of discussing
in detail all of the inherent uncertainties in locating stars in the HR
diagram (Figure 2), and in inference of stellar ages and masses from those diagrams, 
I will simply assume fiducial cluster ages based on
the median apparent age of stars in the mass range 0.3-1.0 M$_\odot$.
With both a disk diagnostic and a method of cluster age estimation we
can now explore the evidence for disk evolution.

Our best effort at empirically measuring the time scale for the
evolution of inner circumstellar accretion disks is represented in Figure 3, 
produced from a sample of $\sim$3000 stars 
located $\sim$50-500 pc from the Sun.   To be included in the sample each
star was required to have the spectral type, optical photometry, and
infrared photometry necessary for calculation of $\Delta(H-K)$ or $\Delta(K-L)$,
as described above.  It should be noted that there are far fewer stars 
with available L-band photometry than available (J)HK photometry.
There are several important points made by these example plots. 

First, even at the earliest evolutionary stages at which stars can be located in 
the HR diagram, the optically thick inner disk fraction does not approach unity. 
There are several well-known examples of objects near the stellar birthline
without any evidence for disks.
This may be influenced by selection effects in that protostars and objects
in transition from the protostellar to the optically revealed stage 
generally lack the spectroscopic data required for inclusion in our sample. 
However, the result is more apparent in the H-K excess figure than in the
K-L excess figure.  If a real effect (as opposed to an effect introduced by
bias in the samples selected for L-band photometry), this indicates strongly
that {\it some} disk evolution does happen very early on for {\it some} stars,
before they become optically visible. 

Second, beyond 1 Myr of age existing samples are less biased by complications
of extinction and self-embeddedness, and hence more representative of underlying 
stellar populations as a whole (if not close to complete for most of the regions 
plotted).  At these ages,
there is a steady decline with time in the fraction of stars showing 
near-infrared excess emission (i.e. optically thick inner disks), 
as well as large scatter at any given age.
We will return to the issue of
the scatter later.  The conversion of diagrams like Figure 3 into 
a frequency distributions of accretion disk lifetimes is the next step,
and really what we want to know rather than disk frequency with age; this
analysis is presented in Hillenbrand, Meyer, \& Carpenter (2006). 

Third, the median lifetime of inner optically thick accretion disks based 
on assessment of modern data may be as short as 2-3 Myr with essentially
no evidence for HKL excess present {\it in the median star} beyond 5 Myr.  
Clearly there are exceptions such as the noted cases of $\sim$10 Myr old
accretion disks.

Other discussions of inner disk lifetimes have used different techniques and
more limited samples of stars (e.g. Walter et al. 1988, Strom et al. 1989, 
Skrutskie et al. 1990, Beckwith et al. 1990, Strom 1995; Haisch et al. 2001).
As with most scientific inquiries, the results derived depend on the 
details of both the samples and the analysis. Within the proposed 
random and systematic uncertainties, all of the above studies are comparable
in their results.
Previous general conclusions regarding inner disk lifetimes in the 3-10
Myr age range are, broadly speaking, similar to our findings of $<$2-3 Myr
for the evolution of the mean disk.  Further, although most disks 
appear to evolve relatively rapidly, a small percentage appear to
retain proto-planetary nebular material for factors of 5-10 longer 
than does the average disk.

\
\subsection{Mid-Disk Dissipation}

\begin{figure}
\hskip 1.5truein
\includegraphics[width=0.45\textwidth,clip]{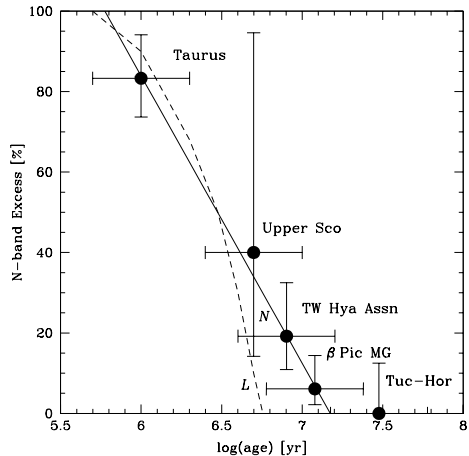}
\caption{
  {\it Terrestrial zone disk fraction vs. stellar age} inferred from
  N-band excess measurements for $\sim$50 stars, taken from 
  Mamajek et al 2004. 
  }
\end{figure}

As emphasized above, near-infrared wavelengths measure hot dust in the innermost
disk regions, the presence of  which is well correlated empirically 
with independent (spectroscopic) evidence for accretion onto the star.
Because only a small amount of dust is required to make the inner disk
optically thick, near-infrared continuum excesses tell us little about
the bulk of the disk mass or surface area, which radiates at much cooler
temperatures and hence longer wavelengths.  Further, because the dynamical
time is a function of radius in the disk, there is some expectation in the
scenario that disk dissipation involves sticking collisions that eventually
generate planetesimals, for disks to evolve in the inner regions first 
and the outer regions later (e.g. Hollenbach et al 2000).  
Thus studying disk frequency with age (or
better yet, disk lifetime) as a function of disk radius is of great interest.

Mid-infrared wavelengths, $\sim$10-90 $\mu$m,   
probe disk radii $\sim$1-5 AU, equivalent to 
the outer terrestrial and inner gas giant planetary zones of our solar system.
To date, observational sensitivity has been the primary hindrance to 
measurement of disk evolution at these wavelengths.  The sensitivity
required at mid-infrared wavelengths  is in fact orders of  magnitude 
more in flux density units than that needed in the near-infrared 
due to the Rayleigh-Jeans fall-off of the stellar photosphere.
Despite the large number of non-detections or upper limits,
previous mid-infrared observations of small samples of young stars
have revealed evolutionary trends.

The most recent statistical results using ground-based equipment
(e.g. Mamajek et al 2004; Metchev et al 2004), when 
considered in the same excess fraction format as Figure 3, show similar 
morphology with $\sim$10 Myr needed for
depletion of 90\% of optically thick terrestrial zone dust.  Figure 4 is reproduced
from Mamajek et al.  2004. The implication is that the terrestrial zone disk dissipation times are perhaps 
consistent with, or at most factors of a few longer than, inner disk dissipation times.  
If true, the combined near- and mid-infrared results suggest that disk evolution is both rapid 
and relatively independent of radius.   However, as was true in the analysis of
inner disk lifetimes, a decreasing fractional excess that is never unity is suggestive of a dispersion
in disk lifetimes, in this case over an order of magnitude in age.

The Spitzer Telescope offers dramatic improvement to heretofore available
mid-infrared continuum excess probes of dust evolution.  Spitzer is sensitive to 
nearby {\it stellar photospheres between 3.5 and at least 24 $\mu$m} with additional
sensitive capability out to 70 $\mu$m.  Spitzer thus enables meaningful statistical studies 
of primordial (and debris - see Meyer, this volume) 
disk evolution within and beyond the terrestrial planet zone.
Advances over the previous IRAS/ISO and ground capability are already 
revolutionizing the field.  Restricting the discussion to only 8 $\mu$m results,
 Silverstone et al. (2006) study both field stars and cluster / association
members $<$30 Myr old, Young et al (2004) present results for a single 30 Myr old cluster
NGC 2547, and  Stauffer et al (2005) discuss the 120 Myr old Pleiades cluster.  All of these
papers reaffirm the basic Mamajek findings that terrestrial zone dust is depleted within 10 Myr,
but add needed statistics.  Further Spitzer results are forthcoming.

At longer wavelengths, 25-60 $\mu$m, data from the IRAS and ISO 
satellites were even more limited in addressing disk evolution problems, again
due to the sensitivity requirements of such investigations.  These platforms 
were not capable of detecting 
the stellar photospheres of young stars at the necessary 150 pc distance.
However, some results at 60 $\mu$m have  been presented 
in the same form as Figure 3 (e.g. Meyer \& Beckwith 2000;
Robberto et al. 1999),   again suggesting consistency with the Mamajek et al results at 10 $\mu$m.
However, Spangler et al. (2001) and  Habing et al. (2001) argue, based on ISO data, for
a much longer mid-infrared disk dissipation time scale, on the order of hundreds of Myr.  
There may be some confusion in these two studies between primordial and debris disks
as a single, continuous evolutionary path is not expected over this long time scale
(see, Decin et al. 2003 for a critical assessment).  

Again, Spitzer will revolutionize the field
due to its increased sensitivity and spatial resolution over previous capabilities.
Results at 24  $\mu$m for the 5 Myr old Upper Sco association 
(Chen et al. 2005) and for the 10 Myr TW Hya association  (Low et al. 2005) 
have appeared thus far. However, the mix of spectral types in these early studies 
relative to the roughly solar-type stars discussed above
make rigorous comparisons of the disk dissipation statistics with radius premature.

\subsection{Outer Disk Dissipation}

Moving outward in wavelength and hence downward in temperature,
millimeter wavelength emission probes the cold outer ($\sim$50-100 AU) disk
regions and is optically thin.  Most millimeter observations (e.g. Andrews \& Williams 2005)
have been directed towards stars younger than $\sim$10$^7$ year, but because of the
distance of these populations, generally place only upper limits on dust 
masses beyond the youngest  phase (e.g. Duvert et al. 2000). 
Recent application of a clever technique to push below formal detection
limits has resulted in more stringent constraints on the typical disk masses 
in several very young clusters (Carpenter 2002; Eisner \& Carpenter 2003), 
finding mean dust masses of $5 \times 10^{-5}$ M$_\odot$ 
(which can be augmented by an assumed gas-to-dust ratio to infer a total mass).
Because of the more dispersed nature of older stars,
there remain few such stringent constraints on typical dust masses 
in the 3-10 Myr old age range.

Dust mass surveys of older (10$^7$ - 10$^9$ year), closer, candidate
{\it debris} disk stars (e.g. Zuckerman \& Becklin 1993, Jewitt 1994, 
Wyatt et al. 2003, Carpenter et al. 2005) also reveal mostly upper limits due
to current sensitivity challenges, but also several detections of very
proximate stars with dust masses as low as 10$^{-8}$ M$_\odot$.
In an analysis of the ensemble of upper limits,
Carpenter et al. (2005) find marginal evidence for continuous evolution in
the dust masses at expected primordial disk ages, from the 1-2 Myr 
young clusters to the 3-10 and 10-30 Myr field stars, which may in fact 
have already transitioned from primordial to debris disks. 

Assessment of primordial disk evolution at radii of several tens to hundreds of AU,
where the bulk of the disk mass resides, thus awaits dramatically improved 
millimeter and sub-millimeter sensitivity.  Such is
on the horizon with the comissioning of CARMA and ALMA.

\section{How Long  Does the Dissipation Process Take, Once Initiated?}

Once the process of disk dissipation starts, how long does it take
for an individual object to transition from optically thick to optically thin
dust?
The expectation is for a short transition, based on calculations of initial grain
growth via pairwise sticking collisions followed by runaway growth that forms
large planetesimals (Moon-sized) on times scales of only $\sim10^5$ yr
(e.g. Wetherill \& Stewart 1993; Weidenshilling \& Cuzzi 1993).
Is there a radial dependence to the disk clearing or do inner, mid, and outer
disk regimes dissipate simultaneously?
While there are clear decreasing trends with advancing stellar age
both in the  fraction of objects exhibiting infrared excess and in the
mean magnitude of the infrared excess (not addressed in the discussion above), 
this does not inform us about the
disk dissipation time for an individual object.   The observed trends and
their dispersion can be used, however, to construct statistical arguments that address
the  duration of the disk dissipation process, as a function of radius.

Historically, a relatively short, less than a few hundred thousand year time 
scale, has been inferred for the transition from an optically thick circumstellar
disk to an optically thin circumstellar disk.  The logic is based on two arguments,
first the disk statistics in binary pairs, and second the small number
(and therefore fraction) of objects found in transition between the optically thick
and optically thick stages.  Binary pairs, particularly in Taurus, have been well
characterized in terms of the well-known CTTS (disked) and WTTS (disk-less) categories.
Numerous studies (e.g. Hartigan, Strom, \& Strom 1994, Prato \& Simon 1997,
Duchene et al. 1999, Hartigan \& Kenyon 2003) have found that the vast majority,
$>$80\%, of binary pairs are either both CTTS or both WTTS with mixed pairs
relatively rare.  This argues that the disk dissipation time is shorter than the absolute
age difference between the members of stellar binaries.

Concerning transition objects, in the well-studied Taurus star-forming region,
for example, V819 Tau and V773 Tau were argued by Skrutskie et al (1990)
to be the only two members out of approximately 150 known found with little or no near-infrared
excess but small mid-infrared excess
\footnote{see Duchene et al 2003 for evidence concerning the multiplicity of V773 Tau
and argument that the apparent excess can be attributed to one of the companions 
rather than betraying a ``fossil" disk.}, a result confirmed by 
Simon \& Prato (1995) and Wolk \& Walter (1996).
This argument relies on the assumption of cluster coevality.
As discussed above, this may not be a valid assumption 
at the few (2-3) Myr level.
Spitzer data presented by Hartmann et al (2005) appear to add several other
objects to the ``transition" category, 
such as CIDA 8, CIDA 11, CIDA 12, CIDA 14, DH Tau, DK TauB, and FP Tau.   
 
Yet other Taurus objects have no evidence for excess out to 10 $\mu$m but
substantial excess at longer wavelengths.  These are different from the sources
detected with excess at or short-wards of 10 $\mu$m,  but in transition from having 
optically thick to optically thin inner disks.  They may be even slightly more
evolved (in a circumstellar sense).  One interpretation is that  
on the time scale that inner disk clearing has
completed, these disks may be transitioning 
from optically thick to optically thin in the mid or outer disk regions.  
GM Aur has long been appreciated
in this category (e.g. Koerner et al. 1993; Rice et al 2003).   
Others with excesses only at long wavelengths were not
detectable with the sensitivity of IRAS but are being revealed by Spitzer, 
for example CoKu Tau4 (D'Alessio et al 2005) and DM Tau (Calvet et al 2005b).

Collectively, both the optically thin and the inner cleared disks can be referred to
as ``transitional."
In regions other than Taurus, the case for transitional disks has also been made.
For example, Gauvin \& Strom (1992) highlighted CS Cha  in Chamaeleon as 
having a large inner cleared region (tens of AU) but a substantial far-infrared excess
indicative of a robust outer disk.  Nordh et al. (1996) 
show 7-15 $\mu$m flux ratios in Chamaeleon that are scattered around {\it either} the colors 
expected from flat/flared disks, {\it or} around photospheric
colors, with essentially no objects located in between these groupings.
These observations support the rapid transition time scales argued for Taurus members.
Low et al (2005)  observed the same effect at longer wavelengths, 24 $\mu$m,
 in the much older TW Hya association.

In summary, in young ($<$3 Myr) star forming regions transition disks rare, with most stars having
circumstellar material that is either consistent with an optically thick disk or not apparent
at all.  Further, there are specific examples of stars with dust in the
terrestrial planet zone (0.1-3 AU) but not in the very inner disk ($<0.05$ AU).  This
suggests that material closest to the star may disappear first, as accretion subsides,
and that the disk is cleared from the inside out. 
In slightly older (10 Myr) regions the only disks left appear to be those in transition,
already evolved or fully cleared in the inner disk regions but retaining mid-infrared excesses
indicative of mid-range disks.  

How does disk clearing occur?
While  photometric studies at infrared and millimeter wavelengths such as those discussed
above can provide statistics
for assessing the dust disk dissipation time scale, and hence the dust disk
lifetime, they tell us very little about the physics of the process.   Studies of
evolutionary changes in the disk structure or dust grain processes, by contrast, do provide
physical insight but are restricted to much smaller samples which can be studied in detail.  
Spectral energy distributions and mineralogy are two tools that can provide insight. 

Typically, grain growth and disk evolution arguments have been made
from measurement of the frequency dependence of continuum
opacity in the expression $\tau_{\nu}(r) = \kappa_{\nu} \times \Sigma (r)$ where
$\kappa_{\nu} \propto \nu^\beta$.  The $\beta=2$ appropriate for
interstellar dust often yields in measurements of optically thin 
sub-/millimeter spectral energy distributions to 
$\beta = 0-1$ (see Miyake \& Nakagawa, 1993).   The effects on the overall
spectral energy distribution of grain growth are presented in a parameter study of
disk geometry and grain properties by D'Alessio et al. (2001).

Detailed spectral energy distributions are most useful when combined 
with spatially resolved imaging at one or more wavelengths, enabling
degeneracies in model parameters to be removed.  Modelling studies of objects in
different circumstellar evolutionary stages e.g., Class 0, Class I, Class II, 
(perhaps even Class III someday) spectral energy distributions can provide constraints on disk
geometry.  Some examples of such work are the analyses by Wood et al. (1998),
Wolf et al. (2003), Eisner et al. (2006), Kitamura et al. (2002), and Calvet et al (2002).
It should be borne in mind, however, that the connections between circumstellar
and {\it stellar} evolutionary states are not yet clear.

As the dust transitions from optically thick to optically thin, perhaps as a function 
of radius, spectroscopy becomes an especially important tool for assessing
grain size distribution and composition.  
Mineralogical studies reveal information about  dust processing,
for example changes in chemical composition or mean grain size.   
There is evidence already for the growth of grains in young disks to sizes larger 
than are expected based on the assumption that disk grain properties are consistent
with those of interstellar dust.    Direct probes 
of grain growth are spectroscopic studies that are sensitive to the opacity from 
particular species having particular size ranges.
Work in the 8-13 $\mu$m atmospheric  window (e.g. Kessler-Silacci et al. 2005, van Boekel et al. 2005)
is being complemented, improved, and extended by Spitzer studies from 5-40 $\mu$m. 
Especially compelling observations would be those that can obtain spatially
resolved mineralogical information.  Intriguing results in this area have emerged
recently from the VLTI (e.g. van Boekel et al 2004)

\section{Present Assessment of  Dust  Clearing Trends with Radius}

A single sentence summary of the above set of results on inner disks, mid-range disks,
and outer disks is that the often quoted ``10 Myr disk lifetime"  is a gross generalization.  
While there are some clear declining trends at several wavelengths in measured disk 
strength and disk frequency with time, the simple fact that we can
consider the quantity ``disk frequency" implies that at any given age some stars have disks while
others do not, and thus a range in disk evolutionary times.  The dispersion in disk lifetimes
is at least factors of a few if not an order of magnitude.  

   There is a some evidence that disk clearing times
may be shorter in the near-infrared than in the mid-infrared though this conclusion is not strong
at present.  The most conservative statement is that 
dust disk dissipation appears to occur within 3-8 Myr for the vast majority of stars, 
with minor evidence for more rapid time scales at smaller radii.  The 
dissipation time for the mean disk may be $<$2-3 Myr.  
Sensitive observations with Spitzer
of statistically significant samples of young stars spanning an appropriate age range
are needed before such conclusions are robust, however.  Such are beginning to emerge.

It should also be noted that
the methods employed to date for statistical study of disks and disk lifetimes
largely {\it detect} the presence or absence of a disk and do not tell us much
about the detailed disk properties (radial/vertical structure, total mass, composition, etc.).  
This is another area in which the improved sensitivity and the spectroscopic 
capabilities of Spitzer along with the spatially resolved imaging capabilities
of ground-based facilities will improve our understanding, though only for 
selected individual objects.

Finally, we reiterate that a complication in developing our 
empirical understanding of the time scales
and physical processes associated with primordial disk dissipation is that
soon after dusty disk material begins agglomerating to form planetesimals,
the proto-planets likely collide and re-form the dust.  
When does a particular system go from being primordial
(dominated by growth of smaller bodies into larger ones) to debris 
(dominated by destruction of larger bodies into smaller ones which
are then removed from the system via Poynting-Robertson drag and stellar wind 
effects)?  For disks surrounding stars with ages in the 5-15 Myr age range 
there is some ambiguity as to whether they are primordial
or debris disks.  Several prominent examples are TW Hya, 
which is still accreting (Muzerolle et al 2000),  Beta Pic and AU Mic, 
both of which are nearby and spatially resolved,
and new spatially unresolved detections in the 5-15 Myr age range
emerging from Spitzer (e.g. Chen et al. 2005; Low et al. 2005; 
Silverstone et al. 2006).

As in the above discussion of primordial optically thick disks,
spatial resolution is the key element for advances in debris disk studies 
with, for example, the color of scattered light providing critical information 
about the radial distribution of grain sizes (e.g. Metchev et al 2005).
Our main diagnostic for observationally distinguishing primordial from debris 
disks is the presence of gas, discussed in this proceedings
in more detail by Najita.   

\section{Implications for Planet Formation}

The discovery of exo-solar planets more than a decade ago made understanding of the
connections between disks (both primordial and secondary/debris) and planets 
more critical than ever. The near ubiquity of circumstellar dust and gas disks around 
very young stars has been
advocated for decades, but only within the past decade uniformly accepted by
the astronomical community.  The turning point was availability of spatially
resolved images of young gaseous and dusty disks
at millimeter, sub-millimeter, infrared, and optical wavelengths.  
Beyond evidence
for disks, the detailed information provided by 1) such images,  2) spectral energy distributions
sampled over more than four decades in wavelength,
and 3) dust and gas spectroscopy, is increasing our understanding of the initial conditions
for planet formation.
This review has concentrated on dust disk diagnostics.

However, detailed understanding of the processes
of star and planet formation requires  statistical 
assessment of global properties and evolutionary trends,
in addition to study of individual objects. Despite the large amount
of data presently available we are only now beginning to achieve the observational
sensitivity needed to probe the full range of disk conditions.
For the assembly of statistics we still need to rely on traditional photometric and spectroscopic 
techniques rather than well sampled spectral energy distributions plus spatially
resolved imaging at multiple well separated wavelengths, which are available in relatively few cases.

With the statistics available at present, there are constraints on disk dissipation
time scales  though they are limited in 
terms of the detail needed to constrain theories.
Evidence for decreasing trends with age in the disk fraction, 
the mean disk accretion rate, and the mean disk mass are apparent.
There are also signs in individual young disks of evolution from interstellar grain parameters.  
What may be most interesting however, is the \emph {large dispersion about the mean 
at any given age}, in all of these trends.  This in particular speaks to
the frequency distribution of paths for solar system formation and evolution.

By establishing the decay with time of primordial dust via
near- and mid-infrared excess around stars of different mass, 
we will take, over the next decade, the first step in understanding
the possibilities for planetary formation.  Studies to determine the time 
scales for dust disk dissipation should be followed by those aiming to 
similarly quantify time scales for gas disk dissipation.  
Fully constraining the time period
over which the raw materials needed for planetary formation are available means, 
ultimately, following the evolution of disk surface density 
as a function of radius from the central star.  One outstanding problem 
in planning for this kind of statistically robust future is that we do not have
adequate samples of stars in the 5-50 Myr age range, a
critical time in planet formation and early solar system evolution.

Various theories of dust settling, planet formation, and planetary migration within disks
are discussed elsewhere in these proceedings.  The limited constraints from theory 
 are consistent with the equally vague precision 
with which disk lifetimes can be inferred from observations 
of potential planetary systems now in the making.  Thus the interpretation
of observations is not -- yet -- the limiting step in solidifying our understanding
of planet formation.

When, where, and how 
frequently do planets form in circumstellar disks? How do forming
planetary systems evolve dynamically?  What is the range in diversity of stable planetary system 
architectures?    How frequent are 
habitable planets?  How unique is our solar system?  These are fairly sophisticated questions
to be asking, especially so given our only rough knowledge of the planet formation process in our own
solar system.  Meteoritic
evidence concerning survival time of the solar nebula suggests ``several
Myr'' as the relevant evolutionary time scale.  Studies, especially those
concerning extinct radionuclides, support this time span for
initial accretion, differentiation, and core formation
(see e.g. review by Wadhwa \& Russell 2000). 
It should be emphasized that although dispersal of the solar {\it nebula}
may occur quickly, the total duration over which inner planet 
formation was completed in fact approached 30-100 Myr.

An overarching goal of these pursuits is to connect
what is observed elsewhere with the history of our own solar solar system,
and hence enhance our appreciation of the uniqueness -- or lack thereof -- 
of it, our Earth, and in some respects the human circumstance.

\end{document}